\documentclass[journal]{IEEEtran}
\usepackage{rotating}
\usepackage{graphicx}
\usepackage[cmex10]{amsmath}
\usepackage{cite}
\usepackage{amssymb,amsfonts}

\begin{document}

\title{ Chain of Separable Binary Goppa Codes and their
Minimal Distance}
\author{Sergey Bezzateev and Natalia Shekhunova
\thanks{S. Bezzateev and N. Shekhunova are  with the Department
of Information Systems and Security , Saint Petersburg State University of Airspace Instrumentation, Saint-Petersburg,
Russia, e-mail: bsv@aanet.ru, sna@delfa.net}}
 
\markboth{IEEE Transaction on Information Theory, Vol. , No. ,  }
{Shell \MakeLowercase{\textit{et al.}}: Chain of Separable Binary Goppa Codes and their Minimal Distance}
\maketitle

\begin{abstract}
It is shown that subclasses of separable binary Goppa codes , $\Gamma(L,G)$ - codes,
with $L=\{ \alpha \in GF(2^{2l}):G(\alpha )\neq 0 \}$ and special Goppa 
polynomials $G(x)$ can  be presented  as a chain of embedded codes. 
The true minimal distance has been obtained for all codes of the chain.
\end{abstract}
 
 
\IEEEpeerreviewmaketitle
 
\section{Introduction}
 
\IEEEPARstart {A}{ }ny $q$-ary Goppa code $\Gamma(L,G)$ -code can be defined 
by two objects:
Goppa polynomial $G(x)$ where $G(x)$ is a polynomial of a degree $t$ over $GF(q^m)$ and
 location set $L$ where $L=\{\alpha \in GF(q^m):G(\alpha )\neq 0\}$.
 
\newtheorem{definition}{Definition}
\begin{definition}
A $q$-ary vector $a=(a_{1}...a_{n})$ of a length $n$ where $n$ is a
cardinality of the set $L=(\alpha_{1},...,\alpha_{n}),\alpha_{i} \in GF(q^m)$ 
is a codeword of the $\Gamma (L,G)$ Goppa code if
and only if  the following equation is satisfied:
\begin{center}
$\sum\limits_{i=1}^{n}a_{i}\frac{1}{x-\alpha _{i}}\equiv 0 \:mod G(x)$.
\end{center}
 
\end{definition}
 
The parity check matrix of the $\Gamma(L,G)$ -code can be presented in the following form:
\begin{center}
$H=\left[ 
\begin{array}{ccc}
\frac{1}{G(\alpha _{1})} & ... & \frac{1}{G(\alpha _{n})} \\ 
\frac{\alpha _{1}}{G(\alpha _{1})} & ... & \frac{\alpha _{n}}{G(\alpha _{n})} \\ 
. & . & . \\ 
\frac{\alpha _{1}^{t-1}}{G(\alpha _{1})} & ... & \frac{\alpha_{n}^{t-1}}{G(\alpha _{n})}%
\end{array}%
\right].$
\end{center}

\indent It is  known that the  $\Gamma(L,G)$ -code has  following parameters \cite{Goppa1}:

the length of the code is equal to the cordinality $n$ of the location set $L$, $n \leq q^m $

the dimension is $k\geq n-tm$ and 

the true minimal distance is $d\geq t+1$.

The  $\Gamma(L,G)$-code is a binary Goppa code if $q=2$.

The $\Gamma(G,L)$- code  is  a separable Goppa code if all roots of its Goppa polynomial $G$ are different.

The following estimation of the true minimal distance for binary separable (L,G)-codes is valid \cite{Goppa1}:
\begin{center} 
$d\geq 2t+1$.
\end{center}

Binary separable Goppa codes have been studied by many authors.  The  binary separable
 codes with the location set  over $GF( 2^{sl} )$ where $s=2,3,...$ are of the greatest interest.
 
M. Loeloeian and J. Conan  were the first who considered the code of this class.
In 1984 they  presented\cite{Loel1} the best known (55,16,19)- Goppa code with the Goppa polynomial:
\begin{center} 
$G(x)=(x-\alpha ^{9})\dot(x-\alpha ^{12})\dot(x-\alpha ^{30})\dot(x-\alpha ^{34})\dot(x-\alpha ^{42})\dot(x-\alpha ^{43})\dot(x-\alpha ^{50})\dot(x-\alpha ^{54})$
\end{center}
where $\alpha$  is a primitive element of $GF(2^{6}).$

In 1986 we considered \cite{Bezz5} this code as a code from  a subclass of Goppa codes with  the Goppa polynomial: \\
$G(x)=x^{t+1}+V^{t}x^{t}+Vx+1$\\
where $V\in GF(2^{2l})$ , $t=2^l$ , $L\subset GF(2^{2l})$ and $n=2^{2l}-t-1$.\\
We have proved \cite{Bezz5}, \cite{Bezz1} that the dimension of these codes is \\
$k \geq n-2l(t-\frac{3}{2})$.

In 1987 M. Loeloeian and J. Conan \cite{Loel2} considered a subclass of Goppa  codes with the  Goppa polynomial:\\
$G(x)=x^{t}+x$\\
where $t=2^l$ ,$L\subset GF(2^{2l})$ and $n=2^{2l}-t$.\\
They also gave the estimation for  the dimension of these codes:\\
$k \geq n-2l(t-\frac{3}{2})-1.$

The same estimation for  the dimension of these codes  was obtained by A.M.Roseiro, J.I.Hall, J.E.Adney and M.Siegel  in \cite{Roseiro1} by using the kernel of an associated trace map.
In this paper, the subclass of codes with polynomial $G(x)=x^{t}+x$  was called as quadratic trace Goppa codes.

In 1995 we described \cite{Bezz2} the subclass of Goppa codes with the polynomial $G(x)=x^{t-1}+1$  and we have proved that the minimal distance of these codes is equal to their design distance.

In 2001 P. Veron \cite{Veron3} investigated  the structure of  trace Goppa codes and proved that the true dimension for these codes is equal to the  estimation obtained previously:\\
 $k \geq n-2l(t-\frac{3}{2})-1$
 
In 2005  P. Veron \cite{Veron2} proved that the  estimation of the  code dimension for Goppa codes with $G(x)=x^{t+1}+V^{t}x^{t}+Vx+1$ and $G(x)=x^{t-1}+1$ is the true dimension  for the codes from this subclasses \cite{Bezz3}. 

In \cite{Bezz6}  and also in \cite{Bom1} (G. Bommer  and F. Blanchet)  and in \cite{Veron1} (P. Veron) it was proved   that all the  mentioned above codes are  a quasi-cyclic binary Goppa codes.

In 2007 G.Maatouk, A.Shokrollahi and M.Cheraghchi \cite{Maatouk1} tried to prove that the class of codes which  was described in \cite{Bezz2} achieved the GV bound.

In this paper,  we present all codes that  were mentioned above as a chain of embedded codes.
We obtain the true minimal distance for these codes. The rest of the paper is organized as follows. 

Section 2 describes the chain of Goppa codes subclasses. 

Section 3 gives several Lemmas for the true minimal distance for  subfield subcodes:\\  $G_{5}(x)=Cx^{t+1}+A^{t}x^{t}+Ax$,\\ $G_{6}(x)=Rx^{t+1}+V^{t}x^{t}+Vx+1$\\
where $R\in GF(2^{l}),V\in GF(2^{2l})$  and $G_{7}(x)=x^{t+1}+1.$

In Section 4 similar Lemmas are presented for quadratic trace Goppa codes:\\
$G_{2}(x)=A^{t}x^{t}+Ax$ and \\
$G_{3}(x)=A^{t}x^{t}+Ax+C$,\\
 where $A\in GF(2^{2l}) $ and $C\in GF(2^{l})$).

In Section 5  we obtain  the true minimal distance for codes that are not subfield or trace codes: these new codes with a Goppa polynomial\\$G_{4}(x)=A^{t}x^{t}+A^{t-1}x^{t-1}+1$\\ have not been described before.

In Conclussion a table with parameters of codes from the code chain and  table of binary quasi-cyclic codes from this chain are presented.

\section{Code chain}

Let us show how to obtain one code from another and  to create a family of embedded codes, a so-called   chain of codes.
 
\begin{definition}
 
Let matrix $H_{1}^{\ast }$ be a parity check matrix of the binary Goppa code with 
 a location set $L_{1}^{\ast }=\{\alpha _{i_{1}},\alpha_{i_{2}},...,\alpha _{i_{n}}\}$ of different nonzero elements from $ GF(2^{2l})$ such that
  $\alpha _{i_{j}}^{t-1}\neq 1$ 
for all $j=1,..,n$ and Goppa polynomial $G_{1}(x)=x^{t-1}+1,t=2^{l}$ :
 
\begin{center}
$H_{1}^{\ast }=\left[ 
\begin{array}{ccc}
\frac{1}{\alpha _{_{i_{1}}}^{t-1}+1} & ... & \frac{1}{\alpha
_{i_{_{n}}}^{t-1}+1} \\ 
\frac{\alpha _{i_{1}}}{\alpha _{i_{1}}^{t-1}+1} & ... & \frac{\alpha _{i_{n}}%
}{\alpha _{i_{_{n}}}^{t-1}+1} \\ 
. & . & . \\ 
\frac{\alpha _{i_{1}}^{t-1}}{\alpha _{i_{1}}^{t-1}+1} & ... & \frac{\alpha
_{i_{n}}^{t-1}}{\alpha _{i_{_{n}}}^{t-1}+1}%
\end{array}%
\right] .$
\end{center}
\end{definition}
 
\newtheorem{lemma}{Lemma}
\begin{lemma}
A row 
$\left[ 
\begin{array}{ccc}
\frac{1}{\alpha _{i_{1}}(\alpha _{_{i_{1}}}^{t-1}+1)} & ... & \frac{1}{%
\alpha _{i_{_{n}}}(\alpha _{i_{n}}^{t-1}+1)}%
\end{array}%
\right] $ can be represented as a linear combination of corresponding rows from
the matrix $H_{1}^{\ast }$.
 
\begin{IEEEproof}
 
For any $\alpha \in L_{1}^{\ast }$
 
\begin{center}
$\frac{\alpha ^{2^{l-1}-1}}{\alpha ^{t-1}+1}=\frac{\alpha ^{2^{l-1}-1}}{%
\alpha ^{(t-1)}(\alpha ^{-(t-1)}+1)}=\frac{1}{\alpha ^{2^{l-1}}(\alpha
^{-(t-1)}+1)}=\frac{1}{\alpha ^{2^{l-1}}(\alpha ^{t-1}+1)^{2^{l}}}=\left( 
\frac{1}{\alpha (\alpha ^{t-1}+1)^{2}}\right) ^{2^{l-1}} .$
\end{center}
 
Therefore the row 
$\left[ 
\begin{array}{ccc}
\frac{1}{\alpha _{i_{1}}(\alpha _{_{i_{1}}}^{t-1}+1)^{2}} & ... & \frac{1}{%
 
\alpha _{i_{_{n}}}(\alpha _{i_{_{n}}}^{t-1}+1)^{2}}%
\end{array}%
\right] $ 
can be obtained from the row 
$\left[ 
\begin{array}{ccc}
\frac{\alpha _{i_{1}}^{2^{l-1}-1}}{\alpha _{i_{1}}^{t-1}+1} & ... & \frac{%
\alpha _{i_{n}}^{2^{l-1}-1}}{\alpha _{i_{_{n}}}^{t-1}+1}%
\end{array}%
\right] $ 
of the matrix $H_{1}^{\ast }$ .
 
For any $\alpha \in L_{1}^{\ast }$
 
\begin{center}
$\frac{1}{\alpha (\alpha ^{t-1}+1)^{2}}+\left( \frac{\alpha ^{2^{l-1}-1}}{%
(\alpha ^{t-1}+1)}\right) ^{2}=\frac{1}{\alpha (\alpha ^{t-1}+1)^{2}}+\frac{%
\alpha ^{2^{l}-2}}{(\alpha ^{t-1}+1)^{2}}=\frac{1}{\alpha (\alpha
^{t-1}+1)^{2}}+\frac{\alpha ^{2^{l}-1}}{\alpha (\alpha ^{t-1}+1)^{2}}=\frac{1%
}{\alpha (\alpha ^{t-1}+1)}$ .
\end{center}
 
Therefore the row 
$\left[ 
\begin{array}{ccc}
\frac{1}{\alpha _{i_{1}}(\alpha _{_{i_{1}}}^{t-1}+1)} & ... & \frac{1}{%
\alpha _{i_{_{n}}}(\alpha _{i_{_{n}}}^{t-1}+1)}%
\end{array}%
\right] $ 
can be obtained from the row 
$\left[ 
\begin{array}{ccc}
\frac{\alpha _{i_{1}}^{2^{l-1}-1}}{\alpha _{i_{1}}^{t-1}+1} & ... & \frac{%
\alpha _{i_{n}}^{2^{l-1}-1}}{\alpha _{i_{_{n}}}^{t-1}+1}%
\end{array}%
\right] $ 
of the matrix $H_{1}^{\ast }$.
\end{IEEEproof}
\end{lemma}
 
\newtheorem{corollary}{Corollary}
\begin{corollary}
 
By using the result of \emph{Lemma 1} we can rewrite the matrix $H_{1}^{\ast }$ 
in the following form:
 
\begin{center}
$H_{1}^{\ast }=\left[ 
\begin{array}{ccc}
\frac{1}{\alpha _{_{i_{1}}}\left( \alpha _{_{i_{1}}}^{t-1}+1\right) } & ...
& \frac{1}{\alpha _{i_{_{n}}}\left( \alpha _{i_{_{n}}}^{t-1}+1\right) } \\ 
\frac{1}{\alpha _{_{i_{1}}}^{t-1}+1} & ... & \frac{1}{\alpha
_{i_{_{n}}}^{t-1}+1} \\ 
\frac{\alpha _{i_{1}}}{\alpha _{i_{1}}^{t-1}+1} & ... & \frac{\alpha _{i_{n}}%
}{\alpha _{i_{_{n}}}^{t-1}+1} \\ 
. & . & . \\ 
\frac{\alpha _{i_{1}}^{t-1}}{\alpha _{i_{1}}^{t-1}+1} & ... & \frac{\alpha
_{i_{n}}^{t-1}}{\alpha _{i_{_{n}}}^{t-1}+1}%
\end{array}%
\right].$
\end{center}
\end{corollary}
 
Now let us obtain the parity check matrix $H_{1}$ for a code $\Gamma (L_{1},G_{1})$
with $G_{1}(x)=x^{t-1}+1,\ t=2^{l}$ and $L_{1}=\{\alpha _{1},\alpha
_{2},...,\alpha _{n_{1}-1},0\},\ n_{1}=2^{2l}-2^{l}+1:$
 
\begin{center}
$H_{1}=\left[ 
\begin{array}{cccc}
\frac{1}{\alpha _{1}^{t-1}+1} & ... & \frac{1}{\alpha _{n_{1}-1}^{t-1}+1} & 1
\\ 
\frac{\alpha _{1}}{\alpha _{1}^{t-1}+1} & ... & \frac{\alpha _{n_{1}-1}}{%
\alpha _{n_{1}-1}^{t-1}+1} & 0 \\ 
. & . & . & . \\ 
\frac{\alpha _{1}^{t-1}}{\alpha _{1}^{t-1}+1} & ... & \frac{\alpha
_{n_{1}-1}^{t-1}}{\alpha _{n_{1}-1}^{t-1}+1} & 0%
\end{array}%
\right].$
\end{center}
 
By using the power $2^{l}$ of the first row from $H_{1}$ we obtain:
 
\begin{center}
$\left[ 
\begin{array}{cccc}
\frac{\alpha _{1}^{t-1}}{\alpha _{1}^{t-1}+1} & ... & \frac{\alpha
_{n_{1}-1}^{t-1}}{\alpha _{n_{1}-1}^{t-1}+1} & 1%
\end{array}%
\right].$
\end{center}
 
The sum of this row and the first row from the matrix $H_{1}$ gives us
 
\begin{center}
$\left[ 
\begin{array}{cccc}
\frac{\alpha _{1}^{t-1}}{\alpha _{1}^{t-1}+1} & ... & \frac{\alpha
_{n_{1}-1}^{t-1}}{\alpha _{n_{1}-1}^{t-1}+1} & 1%
\end{array}%
\right] +
\left[ 
\begin{array}{cccc}
\frac{1}{\alpha _{1}^{t-1}+1} & ... & \frac{1}{\alpha _{n_{1}-1}^{t-1}+1} & 1%
\end{array}%
\right]
=
\left[ 
\begin{array}{cccc}
1 & ... & 1 & 0%
\end{array}%
\right].$
\end{center}
 
Therefore the parity check matrix $H_{1}$ can be rewritten:
 
\begin{center}
$H_{1}=\left[ 
\begin{array}{cccc}
\frac{1}{\alpha _{1}^{t-1}+1} & ... & \frac{1}{\alpha _{n_{1}-1}^{t-1}+1} & 1
\\ 
\frac{\alpha _{1}}{\alpha _{1}^{t-1}+1} & ... & \frac{\alpha _{n_{1}-1}}{%
\alpha _{n_{1}-1}^{t-1}+1} & 0 \\ 
. & . & . & . \\ 
\frac{\alpha _{1}^{t-1}}{\alpha _{1}^{t-1}+1} & ... & \frac{\alpha
_{n_{1}-1}^{t-1}}{\alpha _{n_{1}-1}^{t-1}+1} & 0 \\ 
1 & ... & 1 & 0%
\end{array}%
\right].$
\end{center}
 
Let us define a parity check matrix for a subcode $\Gamma (L_{1}^{\ast},G_{1})$ 
of the code $\Gamma (L_{1},G_{1})$ ,\\ 
$\qquad L_{1}^{\ast}=L_{1}\backslash \{0\}=\{\alpha _{1},\alpha _{2},...,\alpha _{n_{1}-1}\}$,
$n_{1}^{\ast }=n_{1}-1.$
 
\begin{center}
$\left[ 
\begin{array}{ccc}
\frac{1}{\alpha _{1}^{t-1}+1} & ... & \frac{1}{\alpha _{n_{1}-1}^{t-1}+1} \\ 
\frac{\alpha _{1}}{\alpha _{1}^{t-1}+1} & ... & \frac{\alpha _{n_{1}-1}}{%
\alpha _{n_{1}-1}^{t-1}+1} \\ 
. & . & . \\ 
\frac{\alpha _{1}^{t-1}}{\alpha _{1}^{t-1}+1} & ... & \frac{\alpha
_{n_{1}-1}^{t-1}}{\alpha _{n_{1}-1}^{t-1}+1} \\ 
1 & ... & 1%
\end{array}%
\right] =\left[ 
\begin{array}{c}
H_{1}^{\ast } \\ 
1...1%
\end{array}%
\right].$
\end{center}
 
\begin{lemma}
 
$\Gamma (L_{1}^{\ast },G_{1})\equiv \Gamma (L_{2},G_{2})$ where 
$G_{1}(x)=x^{t-1}+1$ and $G_{2}(x)=A^{t}x^{t}+Ax,\ t=2^{l},$ $A$ $\in GF(2^{2l})$.
 
\begin{IEEEproof}
 
Obviously,  $\Gamma (L_{1},G_{2})\equiv \Gamma (L_{2},G_{2}^{\ast })$
where 
$G_{2}^{\ast }(x)=x^{t}+x=x\ast G_{1}(x)$ 
and the Goppa polynomial 
$G_{2}(x)=A^{t}x^{t}+Ax$ 
can be obtained from $G_{2}^{\ast }(x)$ by using the $Ax$ 
substitution for a variable $x$, $A$ $\in GF(2^{2l})$.
 
A parity check matrix $H_{2}$ for the code $\Gamma (L_{2},G_{2}^{\ast })$ with $%
G_{2}^{\ast }(x)=x^{t}+x,\ t=2^{l}$ and 
$L_{2}=\{\alpha _{1},\alpha
_{2},...,\alpha _{n_{2}}\},\ n_{2}=2^{2l}-2^{l}$ is :
 
\begin{center}
$H_{2}=\left[ 
\begin{array}{ccc}
\frac{1}{\alpha _{1}^{t}+\alpha _{1}} & ... & \frac{1}{\alpha
_{n_{2}}^{t}+\alpha _{n_{2}}} \\ 
\frac{\alpha _{1}}{\alpha _{1}^{t}+\alpha _{1}} & ... & \frac{\alpha _{n_{2}}%
}{\alpha _{n_{2}}^{t}+\alpha _{n_{2}}} \\ 
. & . & . \\ 
\frac{\alpha _{1}^{t-1}}{\alpha _{1}^{t}+\alpha _{1}} & ... & \frac{\alpha
_{n_{2}}^{t-1}}{\alpha _{n_{2}}^{t}+\alpha _{n_{2}}}%
\end{array}%
\right].$
\end{center}
 
It is easy to see that this matrix can be rewritten in the following form:

\begin{center}
$H_{2}=\left[ 
\begin{array}{ccc}
\frac{1}{\alpha _{1}\left( \alpha _{1}^{t-1}+1\right) } & ... & \frac{1}{%
\alpha _{n_{2}}\left( \alpha _{n_{2}}^{t-1}+1\right) } \\ 
\frac{1}{\alpha _{1}^{t-1}+1} & ... & \frac{1}{\alpha _{n_{2}}^{t-1}+1} \\ 
. & . & . \\ 
\frac{\alpha _{1}^{t-2}}{\alpha _{1}^{t-1}+1} & ... & \frac{\alpha
_{n_{2}}^{t-2}}{\alpha _{n_{2}}^{t-1}+1}%
\end{array}%
\right] $
\end{center}
 
From  \emph{Corollary 1 } the matrix $\ H_{2}$ is equal to the matrix $H_{1}^{\ast
} $ , therefore 
$\Gamma (L_{1}^{\ast },G_{1})\equiv \Gamma (L_{2},G_{2}^{\ast})\equiv \Gamma (L_{2},G_{2}).$
\end{IEEEproof}
\end{lemma}
(This statement has been proved by P.Veron in \cite{Veron1},\cite{Veron2} by using another approach.)
 
\begin{lemma}
 
$\Gamma (L_{2},G_{2})\equiv \Gamma (L_{3},G_{3})$ where $%
G_{3}(x)=A^{t}x^{t}+Ax+C$, $C\in GF(2^{l})$ and $A$ $\in GF(2^{2l})$.
 
\begin{IEEEproof}
 
Using the $x+\beta $ substitution for a variable $x$ where $\beta \in GF(2^{2l})$
and $\beta \neq A^{-(t-1)}$ we obtain:
 
$G_{2}(x+\beta )=A^{t}(x+\beta )^{t}+A(x+\beta )=A^{t}x^{t}+Ax+(A^{t}\beta ^{t}+A\beta )$\\ 
 where $(A^{t}\beta ^{t}+A\beta)^{2^{l}}=(A\beta +A^{t}\beta ^{t})$
follows from the conditions of \emph{Lemma 3}.
Therefore $C=A^{t}\beta ^{t}+A\beta $ , $C$ $\in GF(2^{l})$ and 
$G_{2}(x+\beta )=A^{t}x^{t}+Ax+C=G_{3}(x)$ .
\end{IEEEproof}
\end{lemma}
 
\begin{lemma}
 
All codewords of the code $\Gamma (L_{4},G_{4})$ with 
$G_{4}(x)=A^{t}x^{t}+A^{t-1}x^{t-1}+1$ 
have the zero value on the position correspondig to the element $0$ from
 $L_{4}=\{\alpha _{1},\alpha _{2},...,\alpha_{n_{4}-1},0\}$.
 
\begin{IEEEproof}
 
Obviously, by substituting $x$ to $A^{-1}x$ in $G_{4}(x)$ we obtain the
same $\Gamma (L_{4},G_{4})$ code with a more simple Goppa polynomial:
$G_{4}(x)=x^{t}+x^{t-1}+1.$
 
Let us consider the parity check matrix of this code:
 
\begin{center}
$H_{4}=\left[ 
\begin{array}{cccc}
\frac{1}{\alpha _{1}^{t}+\alpha _{1}^{t-1}+1} & ... & \frac{1}{\alpha
_{n_{4}-1}^{t}+\alpha _{n_{4}-1}^{t-1}+1} & 1 \\ 
\frac{\alpha _{1}}{\alpha _{1}^{t}+\alpha _{1}^{t-1}+1} & ... & \frac{\alpha
_{n_{4}-1}}{\alpha _{n_{4}-1}^{t}+\alpha _{n_{4}-1}^{t-1}+1} & 0 \\ 
. & . & . & . \\ 
\frac{\alpha _{1}^{t-1}}{\alpha _{1}^{t}+\alpha _{1}^{t-1}+1} & ... & \frac{%
\alpha _{1}^{t-1}}{\alpha _{n_{4}-1}^{t}+\alpha _{n_{4}-1}^{t-1}+1} & 0%
\end{array}%
\right]. $ 
\end{center} 
 
By using the $t-$th degree of the first row of this parity check matrix we can
obtain the following parity check row for $\Gamma (L_{4},G_{4}):$
 
\begin{center}
$r$=$\left[ 
\begin{array}{cccc}
\frac{\alpha _{1}^{t-1}}{\alpha _{1}^{t}+\alpha _{1}^{t-1}+1} & ... & \frac{%
\alpha _{1}^{t-1}}{\alpha _{n_{4}-1}^{t}+\alpha _{n_{4}-1}^{t-1}+1} & 1%
\end{array}%
\right] $.
\end{center}
 
For this row and for the last row of the parity check matrix $H_{4}$ and parity check row $r$
for any codeword $a=(a_{1}...a_{n_{4}})$ of the code $\Gamma (L_{4},G_{4})$ 
the following expressions are valid:
 
$\sum\limits_{i=1}^{n_{4}-1}a_{i}\frac{\alpha _{i}^{t-1}}{\alpha
_{i}^{t}+\alpha _{i}^{t-1}+1}=0$ \ and $\sum\limits_{i=1}^{n_{4}-1}a_{i}%
\frac{\alpha _{i}^{t-1}}{\alpha _{i}^{t}+\alpha _{i}^{t-1}+1}=a_{n_{4}}.$
 
It is possible only in case when $a_{n_{4}}=0$ for all codewords of the code 
$\Gamma (L_{4},G_{4}).$
\end{IEEEproof}
\end{lemma}
 
\begin{corollary}
The code $\Gamma (L_{4},G_{4})$ is equal to the code 
$\Gamma (L_{4}^{\ast},G_{4})$ 
with 
$n_{4}^{\ast }=n_{4}-1$, $k_{4}^{\ast }=k_{4}$
 and 
$L_{4}^{\ast }=\{\alpha_{1},\alpha _{2},...,\alpha _{n_{4}-1}\}$.
 
The parity check matrix $H_{4}^{\ast }$of the code $\Gamma (L_{4}^{\ast },G_{4})$ is:
 
\begin{center}
$H_{4}^{\ast }=\left[ 
\begin{array}{ccc}
\frac{1}{\alpha _{1}^{t}+\alpha _{1}^{t-1}+1} & ... & \frac{1}{\alpha
_{n_{4}-1}^{t}+\alpha _{n_{4}-1}^{t-1}+1} \\ 
\frac{\alpha _{1}}{\alpha _{1}^{t}+\alpha _{1}^{t-1}+1} & ... & \frac{\alpha
_{n_{4}-1}}{\alpha _{n_{4}-1}^{t}+\alpha _{n_{4}-1}^{t-1}+1} \\ 
. & . & . \\ 
\frac{\alpha _{1}^{t-1}}{\alpha _{1}^{t}+\alpha _{1}^{t-1}+1} & ... & \frac{%
\alpha _{1}^{t-1}}{\alpha _{n_{4}-1}^{t}+\alpha _{n_{4}-1}^{t-1}+1}%
\end{array}%
\right] $.
\end{center}
\end{corollary}
 
\begin{lemma}
 
A row 
$\left[ 
\begin{array}{ccc}
\frac{1}{\alpha _{i_{1}}(\alpha _{1}^{t}+\alpha _{1}^{t-1}+1)} & ... & \frac{%
1}{\alpha _{i_{_{n}}}(\alpha _{1}^{t}+\alpha _{1}^{t-1}+1)}%
\end{array}%
\right] $ 
can be represented as  a linear combination of the corresponding rows from
the matrix $H_{4}^{\ast }$.
 
\begin{IEEEproof}
 
For any $\alpha \in L_{4}^{\ast }$
 
\begin{center}
$\frac{\alpha ^{2^{l-1}-1}}{(\alpha ^{t}+\alpha ^{t-1}+1)}=\frac{\alpha
^{2^{l-1}-1}}{\alpha ^{(t-1)}(\alpha +\alpha ^{-(t-1)}+1)}=\frac{1}{\alpha
^{2^{l-1}}(\alpha +\alpha ^{-(t-1)}+1)}=\frac{1}{\alpha ^{2^{l-1}}(\alpha
^{t}+\alpha ^{t-1}+1)^{2^{l}}}=\left( \frac{1}{\alpha (\alpha ^{t}+\alpha
^{t-1}+1)^{2}}\right) ^{2^{l-1}}$.
\end{center}
 
Therefore a row 
\begin{center}
$\left[ 
\begin{array}{ccc}
\frac{1}{\alpha _{1}(\alpha _{1}^{t}+\alpha _{1}^{t-1}+1)^{2}} & ... & \frac{%
1}{\alpha _{n_{4}}(\alpha _{n_{4}-1}^{t}+\alpha _{n_{4}-1}^{t-1}+1)^{2}}%
\end{array}%
\right] $ 
\end{center}
can be obtained from the row 
\begin{center}
$\left[ 
\begin{array}{ccc}
\frac{\alpha _{1}^{2^{l-1}-1}}{\alpha _{1}^{t}+\alpha _{1}^{t-1}+1} & ... & 
\frac{\alpha _{n_{4}}^{2^{l-1}-1}}{\alpha _{n_{4}-1}^{t}+\alpha
_{n_{4}-1}^{t-1}+1}%
\end{array}%
\right] $ 
\end{center}
of the matrix $H_{4}^{\ast }$ .
 
For any $\alpha \in L_{4}^{\ast }$
 
\begin{center}
$\frac{1}{\alpha (\alpha ^{t}+\alpha ^{t-1}+1)^{2}}+\left( \frac{\alpha
^{2^{l-1}-1}}{(\alpha ^{t}+\alpha ^{t-1}+1)}\right) ^{2}=\frac{1}{\alpha
(\alpha ^{t}+\alpha ^{t-1}+1)^{2}}+\frac{\alpha ^{2^{l}-2}}{(\alpha
^{t}+\alpha ^{t-1}+1)^{2}}=\frac{1}{\alpha (\alpha ^{t}+\alpha ^{t-1}+1)^{2}}%
+\frac{\alpha ^{2^{l}-1}}{\alpha (\alpha ^{t}+\alpha ^{t-1}+1)^{2}}=\frac{1}{%
\alpha (\alpha ^{t}+\alpha ^{t-1}+1)}+\frac{\alpha ^{2^{l}}}{\alpha (\alpha
^{t}+\alpha ^{t-1}+1)^{2}}$ .
\end{center}
 
For any $\alpha \in L_{4}^{\ast }$
 
$\frac{\alpha ^{2^{l}}}{\alpha (\alpha ^{t}+\alpha ^{t-1}+1)^{2}}=\frac{%
\alpha ^{2^{l}-1}}{(\alpha ^{t}+\alpha ^{t-1}+1)^{2}}=\left( \frac{\alpha
^{2^{l-1}}}{(\alpha ^{t}+\alpha ^{t-1}+1)}\right) ^{2}+\left( \frac{1}{%
(\alpha ^{t}+\alpha ^{t-1}+1)}\right) ^{2}+\frac{1}{(\alpha ^{t}+\alpha
^{t-1}+1)}$ .
 
Therefore a row 
$\left[ 
\begin{array}{ccc}
\frac{1}{\alpha _{1}(\alpha _{1}^{t}+\alpha _{1}^{t-1}+1)} & ... & \frac{1}{%
\alpha _{n_{4}}(\alpha _{n_{4}-1}^{t}+\alpha _{n_{4}-1}^{t-1}+1)}%
\end{array}%
\right] $ 
can be obtained from the  rows 
\begin{center}
$\left[ 
\begin{array}{ccc}
\frac{\alpha _{1}^{2^{l-1}-1}}{\alpha _{1}^{t}+\alpha _{1}^{t-1}+1} & ... & 
\frac{\alpha _{n_{4}}^{2^{l-1}-1}}{\alpha _{n_{4}-1}^{t}+\alpha
_{n_{4}-1}^{t-1}+1}%
\end{array}%
\right] $ 
,
\end{center}
\begin{center}
$\left[ 
\begin{array}{ccc}
\frac{\alpha _{1}^{2^{l-1}}}{\alpha _{1}^{t}+\alpha _{1}^{t-1}+1} & ... & 
\frac{\alpha _{n_{4}}^{2^{l-1}}}{\alpha _{n_{4}-1}^{t}+\alpha
_{n_{4}-1}^{t-1}+1}%
\end{array}%
\right] $
\end{center}
and
\begin{center}
$\left[ 
\begin{array}{ccc}
\frac{1}{\alpha _{1}^{t}+\alpha _{1}^{t-1}+1} & ... & \frac{1}{\alpha
_{n_{4}-1}^{t}+\alpha _{n_{4}-1}^{t-1}+1}%
\end{array}%
\right] $
\end{center}
 of the matrix $H_{4}^{\ast }$.
\end{IEEEproof}
\end{lemma}
 
\begin{corollary}
 
By using the result of \emph{Lemma 5} we can rewrite matrix $H_{4}^{\ast }$ in the
following form:
 
\begin{center}
$H_{4}^{\ast }=\left[ 
\begin{array}{ccc}
\frac{1}{\alpha _{1}(\alpha _{1}^{t}+\alpha _{1}^{t-1}+1)} & ... & \frac{1}{%
\alpha _{n_{4}}(\alpha _{n_{4}-1}^{t}+\alpha _{n_{4}-1}^{t-1}+1)} \\ 
\frac{1}{\alpha _{1}^{t}+\alpha _{1}^{t-1}+1} & ... & \frac{1}{\alpha
_{n_{4}-1}^{t}+\alpha _{n_{4}-1}^{t-1}+1} \\ 
\frac{\alpha _{1}}{\alpha _{1}^{t}+\alpha _{1}^{t-1}+1} & ... & \frac{\alpha
_{n_{4}-1}}{\alpha _{n_{4}-1}^{t}+\alpha _{n_{4}-1}^{t-1}+1} \\ 
. & . & . \\ 
\frac{\alpha _{1}^{t-1}}{\alpha _{1}^{t}+\alpha _{1}^{t-1}+1} & ... & \frac{%
\alpha _{1}^{t-1}}{\alpha _{n_{4}-1}^{t}+\alpha _{n_{4}-1}^{t-1}+1}%
\end{array}%
\right] $ .
\end{center}
\end{corollary}

Now consider a code 
$\Gamma (L_{3},G_{3})$ with $G_{3}(x)=A^{t}x^{t}+Ax+C,\
t=2^{l}$ and $L_{1}=\{\alpha _{1},\alpha _{2},...,\alpha _{n_{3}-1},0\},\
n_{3}=2^{2l}-2^{l}.$ 
Obviously, by substituting $x$ to $A^{-1}Cx$ in $%
G_{3}(x)$ we obtain the same $\Gamma (L_{3},G_{3})$ code with a more simple 
Goppa polynomial $G_{3}(x)=x^{t}+x+1.$ The parity check matrix for this code is:
 
\begin{center}
$H_{3}=\left[ 
\begin{array}{cccc}
\frac{1}{\alpha _{1}^{t}+\alpha _{1}+1} & ... & \frac{1}{\alpha
_{n_{3}-1}^{t}+\alpha _{n_{3}-1}+1} & 1 \\ 
\frac{\alpha _{1}}{\alpha _{1}^{t}+\alpha _{1}+1} & ... & \frac{\alpha
_{n_{3}-1}}{\alpha _{n_{3}-1}^{t}+\alpha _{n_{3}-1}+1} & 0 \\ 
... & ... & ... & 0 \\ 
\frac{\alpha _{1}^{t-1}}{\alpha _{1}^{t}+\alpha _{1}+1} & ... & \frac{\alpha
_{n_{3}-1}^{t-1}}{\alpha _{n_{3}-1}^{t}+\alpha _{n_{3}-1}+1} & 0%
\end{array}%
\right] $ .
\end{center}
 
It is easy to show that if we have a row $r_{1}$ in the parity check matrix $H_{3}$ : 
 
\begin{center}
$r_{1}=\left[ 
\begin{array}{cccccc}
\frac{\alpha _{1}}{\alpha _{1}^{t}+\alpha _{1}+1} & ... & \frac{\alpha _{i}}{%
\alpha _{i}^{t}+\alpha _{i}+1} & ... & \frac{\alpha _{n_{3}-1}}{\alpha
_{n_{3}}^{t}+\alpha _{n_{3}}+1} & 0%
\end{array}%
\right], $
\end{center}
 
then the row
 
\begin{center}
$r_{t}=\left[ 
\begin{array}{cccccc}
\frac{\alpha _{1}^{t}}{\alpha _{1}^{t}+\alpha _{1}+1} & ... & \frac{\alpha
_{i}^{t}}{\alpha _{i}^{t}+\alpha _{i}+1} & ... & \frac{\alpha _{n_{3}-1}^{t}%
}{\alpha _{n_{3}-1}^{t}+\alpha _{n_{3}-1}+1} & 0%
\end{array}%
\right] $
\end{center}
 
is in the parity check matrix of this code, too.
 
Therefore we obtain the following row:
 
\begin{center}
$r^{\ast }=\left[ 
\begin{array}{ccccc}
1 & ... & 1 & ... & 1%
\end{array}%
\right] $
\end{center}
 
from these two rows and the first row of matrix $H_{3}:$
 
\begin{center}
$r_{0}=\left[ 
\begin{array}{cccccc}
\frac{1}{\alpha _{1}^{t}+\alpha _{1}+1} & ... & \frac{1}{\alpha
_{i}^{t}+\alpha _{i}+1} & ... & \frac{1}{\alpha _{n_{3}-1}^{t}+\alpha
_{n_{3}-1}+1} & 1%
\end{array}%
\right]. $
\end{center}

By using the substitution $\alpha _{i}\longrightarrow \alpha _{i}^{-1}$ the matrix 
$H_{3}$ will become:
 
\begin{center}
$H_{3}=\left[ 
\begin{array}{cccc}
\frac{\alpha _{1}^{t-1}}{\alpha _{1}^{t}+\alpha _{1}^{t-1}+1} & ... & \frac{%
\alpha _{n_{3}-1}^{t-1}}{\alpha _{n_{3}-1}^{t}+\alpha _{n_{3}-1}^{t-1}+1} & 1
\\ 
\frac{\alpha _{1}^{i-2}}{\alpha _{1}^{t}+\alpha _{1}^{t-1}+1} & ... & \frac{%
\alpha _{n_{3}-1}^{t-2}}{\alpha _{n_{3}-1}^{t}+\alpha _{n_{3}-1}^{t-1}+1} & 0
\\ 
... & ... & ... & 0 \\ 
\frac{\alpha _{1}}{\alpha _{1}^{t}+\alpha _{1}^{t-1}+1} &  & \frac{\alpha
_{n_{3}-1}}{\alpha _{n_{3}-1}^{t}+\alpha _{n_{3}-1}^{t-1}+1} &  \\ 
\frac{1}{\alpha _{1}^{t}+\alpha _{1}^{t-1}+1} & ... & \frac{1}{\alpha
_{n_{3}-1}^{t}+\alpha _{n_{3}-1}^{t-1}+1} & 0%
\end{array}%
\right]$
\end{center}
 
and the row $r_{t}$\ will be rewritten as:
 
\begin{center}
$r_{t}=\left[ 
\begin{array}{cccccc}
\frac{\alpha _{1}^{t}}{\alpha _{1}^{t}+\alpha _{1}^{t-1}+1} & ... & \frac{%
\alpha _{i}^{t}}{\alpha _{i}^{t}+\alpha _{i}^{t-1}+1} & ... & \frac{\alpha
_{n_{3}-1}^{t}}{\alpha _{n_{3}-1}^{t}+\alpha _{n_{3}-1}^{t-1}+1} & 0%
\end{array}%
\right]$.
\end{center}
 
The first row of the new matrix $H_{3}$ will be the sum of its last
row, the row $r_{t}$ and  row $r^{\ast }.$

For the i-th element of the first row we will obtain:
 
\begin{center}
$\frac{\alpha _{i}^{t-1}}{\alpha _{i}^{t}+\alpha _{i}^{t-1}+1}=\frac{\alpha
_{i}^{t}}{\alpha _{i}^{t}+\alpha _{i}^{t-1}+1}+\frac{1}{\alpha
_{i}^{t}+\alpha _{i}^{t-1}+1}+1$
\end{center}
 
and for the element in the last column of the first row:
 
\begin{center}
$1=0+0+1$.
\end{center}
 
Consequently, it is possible to rewrite the matrix $H_{3}$:
 
\begin{center}
$H_{3}=\left[ 
\begin{array}{cccc}
1 & ... & 1 & 1 \\ 
\frac{\alpha _{1}^{i-1}}{\alpha _{1}^{t}+\alpha _{1}^{t-1}+1} & ... & \frac{%
\alpha _{n_{3}-1}^{t-1}}{\alpha _{n_{3}-1}^{t}+\alpha _{n_{3}-1}^{t-1}+1} & 0
\\ 
\frac{\alpha _{1}^{i-2}}{\alpha _{1}^{t}+\alpha _{1}^{t-1}+1} & ... & \frac{%
\alpha _{n_{3}-1}^{t-2}}{\alpha _{n_{3}-1}^{t}+\alpha _{n_{3}-1}^{t-1}+1} & 0
\\ 
 
... & ... & ... & ... \\ 
\frac{\alpha _{1}}{\alpha _{1}^{t}+\alpha _{1}^{t-1}+1} & ... & \frac{\alpha
_{n_{3}-1}}{\alpha _{n_{3}-1}^{t}+\alpha _{n_{3}-1}^{t-1}+1} & 0 \\ 
\frac{1}{\alpha _{1}^{t}+\alpha _{1}^{t-1}+1} & ... & \frac{1}{\alpha
_{n_{3}-1}^{t}+\alpha _{n_{3}-1}^{t-1}+1} & 0%
\end{array}%
\right] =\left[ 
\begin{array}{cc}
1..1 & 1 \\ 
H_{4}^{\ast } & 0%
\end{array}%
\right].$

\end{center}
  \begin{figure*}[t]
	\center

	\includegraphics[width=0.5\textwidth]{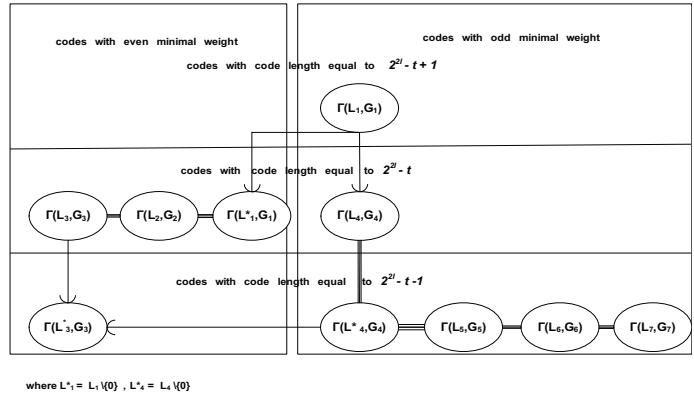}
	\caption{Code chain}
	~\label{figure-1}
\end{figure*}

\begin{definition}
 
Let us define a subcode $\Gamma (L_{3}^{\ast },G_{3})$ of the code $\Gamma
(L_{3},G_{3})$ as shortened by a position corresponding to the element $0$ from 
$L_{3}=\{\alpha _{1},\alpha _{2},...,\alpha _{n_{3}-1},0\}$.  
Hence 
$\Gamma(L_{3}^{\ast },G_{3})\subset \Gamma (L_{3},G_{3})$ and 
$n_{3}^{\ast}=n_{3}-1 $ , $k_{3}^{\ast}=k_{3}-1 $ and $L_{3}^{\ast }=L_{3}\backslash \{0\}$ .
\end{definition}
 
\begin{lemma}
 
$\Gamma (L_{3}^{\ast },G_{3})\subset \Gamma (L_{4},G_{4})$ where $%
G_{4}(x)=A^{t}x^{t}+A^{t-1}x^{t-1}+1$, $A$ $\in GF(2^{2l})$.
 
\begin{IEEEproof}
 
It follows directly  from the above presentation of the matrix $H_{3}.$
\end{IEEEproof}
\end{lemma}
 
\begin{corollary}
 
$\left[ 
\begin{array}{c}
H_{1}^{\ast } \\ 
\begin{array}{cccc}
1 & 1 & ... & 1%
\end{array}%
\end{array}%
\right]$=
$\left[ 
\begin{array}{cc}
H_{4}^{\ast } & 0 \\ 
1...1 & 1%
\end{array}
\right].$
 
\begin{IEEEproof}
 
It follows directly from Lemma 3 where we have proved  the equivalence of two codes: $%
\Gamma (L_{2},G_{2})$ and$\ \Gamma (L_{3},G_{3})$ .
\end{IEEEproof}
\end{corollary}
\bigskip 
 
\begin{lemma}
 
$\Gamma (L_{4},G_{4})\equiv \Gamma (L_{5},G_{5})$ where $%
G_{5}(x)=Cx^{t+1}+A^{t}x^{t}+Ax$,$\ C\in GF(2^{l}),$ and $A$ $\in GF(2^{2l})$ .
 
\begin{IEEEproof}
 
It is easy to show that $G_{5}(x)=x\ast G_{4}(x)\ $and $L_{5}=L_{4}\backslash \{0\}$.
 
Obviously, by substiting $x$ to $A^{-1}Cx$ in $G_{5}(x)$ we obtain the
same $\Gamma (L_{5},G_{5}^{\ast })$ code with a more simple $G_{5}^{\ast
}(x)=x^{t+1}+x^{t}+x.$ 

The parity-check matrix for this code is:
 
\begin{center}
$H_{5}=\left[ 
\begin{array}{ccc}
\frac{1}{\alpha _{1}^{t+1}+\alpha _{1}^{t}+\alpha _{1}} & ... & \frac{1}{%
\alpha _{n_{5}}^{t+1}+\alpha _{n_{5}}^{t}+\alpha _{n_{5}}} \\ 
\frac{\alpha _{1}}{\alpha _{1}^{t+1}+\alpha _{1}^{t}+\alpha _{1}} & ... & 
\frac{\alpha _{n_{5}}}{\alpha _{n_{5}}^{t+1}+\alpha _{n_{5}}^{t}+\alpha
_{n_{5}}} \\ 
. & . & . \\ 
\frac{\alpha _{1}^{t}}{\alpha _{1}^{t+1}+\alpha _{1}^{t}+\alpha _{1}} & ...
& \frac{\alpha _{n_{5}}^{t}}{\alpha _{n_{4}-1}^{t+1}+\alpha
_{n_{4}-1}^{t}+\alpha _{n_{5}}}%
\end{array}%
\right] =\left[ 
\begin{array}{ccc}
\frac{1}{\alpha _{1}(\alpha _{1}^{t}+\alpha ^{t-1}+1)} & ... & \frac{1}{%
\alpha _{n_{5}}(\alpha _{n_{5}}^{t}+\alpha _{n_{5}}^{t-1}+1)} \\ 
\frac{1}{\alpha _{1}^{t}+\alpha _{1}^{t-1}+\alpha _{1}} & ... & \frac{{}}{%
\alpha _{n_{5}}^{t}+\alpha _{n_{5}}^{t-1}+\alpha _{n_{5}}} \\ 
. & . & . \\ 
\frac{\alpha _{1}^{t-1}}{\alpha _{1}^{t}+\alpha _{1}^{t-1}+\alpha _{1}} & ...
& \frac{\alpha _{n_{5}}^{t-1}}{\alpha _{n_{4}-1}^{t}+\alpha
_{n_{4}-1}^{t-1}+\alpha _{n_{5}}}%
\end{array}%
\right] .$ 
\end{center}
Therefore $H_{5}=H_{4}^{\ast }$ according to \emph{Corollary 2} .
\end{IEEEproof}
\end{lemma}
 
\begin{lemma}
 
$\Gamma (L_{5},G_{5})\equiv \Gamma (L_{6},G_{6})$ where $%
G_{6}(x)=Rx^{t+1}+V^{t}x^{t}+Vx+1$, $R\in GF(2^{l}),$ and 
$V$ $\in GF(2^{2l}) $ .
 
\begin{IEEEproof}
 
Using  the $x+\beta $ substitution for a variable $x$ we obtain:
 
\begin{center}
$G_{5}(x+\beta )=$ $Cx^{t+1}+(A^{t}+C\beta )x^{t}+(A+C\beta ^{t})x+(C\beta
^{t+1}+A^{t}\beta ^{t}+A\beta )$ 
\end{center}
 
where $\beta :$ $\beta \in GF(2^{2l})$ and $\beta \neq \frac{A^{t}}{C}$.
Notice that $(A+C\beta^{t})=(A^{t}+C\beta )^{t}$ and
 
\begin{center}
$(C\beta ^{t+1}+A^{t}\beta ^{t}+A\beta )^{t}=C\beta ^{t+1}+A\beta
+A^{t}\beta ^{t}$.
\end{center}
 
Therefore $(C\beta ^{t+1}+A^{t}\beta ^{t}+A\beta )\in GF(2^{l}).$
 
\begin{center}
$G_{6}(x)=\frac{1}{C\beta ^{t+1}+A^{t}\beta ^{t}+A\beta }G_{5}(x+\beta )=%
\frac{C}{(C\beta ^{t+1}+A^{t}\beta ^{t}+A\beta )}x^{t+1}+\frac{(A^{t}+C\beta
)}{(C\beta ^{t+1}+A^{t}\beta ^{t}+A\beta )}x^{t}+\frac{(A+C\beta ^{t})}{%
(C\beta ^{t+1}+A^{t}\beta ^{t}+A\beta )}x+1,$
 
$G_{6}(x)=Rx^{t+1}+V^{t}x^{t}+Vx+1$
\end{center}
 
where $R=\frac{C}{(C\beta ^{t+1}+A^{t}\beta ^{t}+A\beta )}$ and $V=\frac{%
(A+C\beta ^{t})}{(C\beta ^{t+1}+A^{t}\beta ^{t}+A\beta )}$.
 
This means that $\Gamma (L_{5},G_{5})\equiv \Gamma (L_{6},G_{6})$.
\end{IEEEproof}
\end{lemma}
 
\begin{lemma}
 
$\Gamma (L_{6},G_{6})\equiv \Gamma (L_{7},G_{7})$ where $G_{7}(x)=Bx^{t+1}+1$%
, $B=\alpha ^{2^{l}-1},$ and $\alpha$ is a primitive element of $GF(2^{2l})$.
 
\begin{IEEEproof}
 
It can be proved in  the same way as the previous Lemma by using  the $x+\beta $
substitution for a variable $x$  where $\beta =\frac{V^{t}}{R}$ .
\end{IEEEproof}
\end{lemma}
In Figure~\ref{figure-1} we present the  structure of the code chain. It is possible to define the similar code chain for the codes described in paper \cite{Bezz4}.
 
\section{Minimal distance of subfield subcodes}

\begin{lemma}
 
The minimal distance of the last Goppa code in the chain $\Gamma (L_{7},G_{7})$
exactly equals to its design distance, i.e. $d=2(t+1)+1$.
 
\begin{IEEEproof}
 
It is easy to show that a polynomial $\ x^{2^{l}+1}-1$ can be presented as a
product $\prod\limits_{i=1}^{2^{l}+1}(x-\alpha ^{i(2^{l}-1)})$ where $\alpha 
$ is a primitve element of $GF(2^{2l})$. Choose some element $A$ from $%
GF(2^{2l})$ such that $A^{2^{l}+1}$ $\neq 1$ and let $B=A^{-1}.$ 
Thus we can
calculate two polynomials with all different roots $\{A\alpha
^{i(2^{l}-1)}\} $ and $\{B\alpha ^{i(2^{l}-1)}\}$ $i=1,..,2^{l}+1.$
 
\begin{center}
$x^{2^{l}+1}-A^{2^{l}+1}=\prod\limits_{i=1}^{2^{l}+1}(x-A\alpha
^{i(2^{l}-1)}),$
 
$x^{2^{l}+1}-B^{2^{l}+1}=\prod\limits_{i=1}^{2^{l}+1}(x-B\alpha
^{i(2^{l}-1)})$.
\end{center}
 
The result of the multiplication of these two polynomials and $x$ :
 
\begin{center}
$x (x^{2^{l}+1}-A^{2^{l}+1})
(x^{2^{l}+1}-B^{2^{l}+1})=x^{2^{l+1}+3}-(A^{2^{l}+1}+B^{2^{l}+1})x^{2^{l}+2}-x 
$.
\end{center}
 
A formal derivative of result of this multiplication:
 
\begin{center}
$x^{2^{l+1}+2}-1$.
\end{center}
 
Now consider a binary vector $a=(a_{0}a_{1}...a_{n})$ with nonzero elements
on and only on positions $\beta _{j}$, $(j=1,..,(2^{l+1}+3))$ from the following
subset of $L$:
 
\begin{center}
$\{\{A\alpha ^{i(2^{l}-1)}\},i=1,..,2^{l}+1\}\cup \{\{B\alpha
^{i(2^{l}-1)}\},i=1,..,2^{l}+1\}\cup \{0\}$ .
\end{center}

From the definition of the Goppa code, this vector will be a codeword of $\Gamma
(L_{7},G_{7})\ :$
 
\begin{center}
$\sum\limits_{j=1}^{2^{l+1}+3}a_{i}\frac{1}{x-\beta _{j}}\equiv \frac{%
x^{2^{l+1}+2}-1}{x^{2^{l+1}+3}-(A^{2^{l}+1}+B^{2^{l}+1})x^{2^{l}+2}-x}\equiv
0 mod x^{2^{l}+1}-1$ 
\end{center}
 
where $\beta _{j}\in \{\{A\alpha ^{i(2^{l}-1)}\},i=1,..,2^{l}+1\}\cup
\{\{B\alpha ^{i(2^{l}-1)}\},i=1,..,2^{l}+1\}\cup \{0\}$.
 
Therefore the minimal distance of the Goppa code $\Gamma (L_{7},G_{7})$  is equal to the  design distance $2^{l+1}+3$.
\end{IEEEproof}
\end{lemma}
 
\begin{corollary}
 
The minimal distance of the  equivalent Goppa codes $\Gamma (L_{6},G_{6})$ and $\Gamma (L_{5},G_{5})$  is exactly equal to its design distance, i.e. $d=2(t+1)+1$.
\end{corollary}

\section{Minimal distance of quadratic trace subcodes}
 
\begin{lemma}
 
The minimal distance of the equivalent Goppa codes $\Gamma (L_{2},G_{2})$ and $%
\Gamma (L_{3},G_{3})$ is exactly equal to the minimal even weight of  a codeword of the code $\Gamma (L_{5},G_{5})\equiv \Gamma (L_{6},G_{6})\equiv \Gamma (L_{7},G_{7})$ \ , i.e. $d\geq 2(t+1)+2$.
 
\begin{IEEEproof}
 
It  follows directly from the parity check matrixes 
$H_{3}, H_{2}$ and $H_{5}$.
\end{IEEEproof}
\end{lemma}

It is necessary to note that P.Veron in \cite{Veron1} has proved that the Hamming weight of all codewords of these codes is even.

\section{Minimal distance of the new code}
 
\begin{lemma}
The minimal distance of the $\Gamma (L_{4},G_{4})$  and $\Gamma (L_{4}^{\ast },G_{4})$ Goppa codes is exactly equal to its design distance, i.e. $d=2(t+1)+1$ .
\begin{IEEEproof}
It  follows  directly from  the equivalence of codes $\Gamma (L_{4},G_{4})$ and $\Gamma (L_{5},G_{5})$ (Lemma 7).
\end{IEEEproof}
\end{lemma}

\section{Conclusion}
Parameters of the codes forming a chain are presented in Table 1. 

\begin{figure*}[!p]

\begin{center}
Table 1 \ Parameters of the code chain
\end{center}

\begin{tabular}{|c|c|c|c|}
\hline
\begin{tabular}{l}
$\Gamma $(L, G)-code with \\ 
parity check matrix%
\end{tabular}
& code length & $%
\begin{tabular}{l}
number of \\ 
information symbols%
\end{tabular}%
$ & minimal distance \\ \hline
\begin{tabular}{l}
$\Gamma (L_{1},G_{1})$, where \\ 
$G_{1}(x)=x^{t-1}+1$ \\ 
parity check matrix: $H_{1}$%
\end{tabular}
& $n_{1}=2^{2l}-t+1$ & $k_{1}=2^{2l}-t-2l(t-\frac{3}{2})$\cite{Veron2} & $%
d_{1}=2t-1$\cite{Bezz2} \\ \hline
\begin{tabular}{l}
$\Gamma (L_{1}^{\ast },G_{1})$ , where \\ 
$G_{1}(x)=x^{t-1}+1$ , $L_{1}^{\ast }=L_{1}\backslash \{0\}$ \\ 
parity check matrix:  
$\left[ 
\begin{array}{c}
H_{1}^{\ast } \\ 
1...1%
\end{array}%
\right] $%
\end{tabular}
& $n_{1}^{\ast }=2^{2l}-t$ & $k_{1}^{\ast }=k_{1}-1$ & $%
\begin{tabular}{l}
$d_{1}^{\ast }$ is even \\ 
and equals to \\ 
the minimal \\ 
even weight \\ 
of a codeword \\ 
from the code \\ 
$\Gamma (L_{1},G_{1})$%
\end{tabular}%
$ \\ \hline
\begin{tabular}{l}
$\Gamma (L_{2},G_{2})$, where \\ 
$G_{2}(x)=A^{t}x^{t}+Ax$ \\ 
parity check matrix: 
$\left[ 
\begin{array}{c}
H_{1}^{\ast } \\ 
1...1%
\end{array}%
\right] $%
\end{tabular}
& $n_{2}=2^{2l}-t$ & 
\begin{tabular}{l}
$k_{2}=k_{1}-1$ (Lemma 2) \\ 
$k_{2}=2^{2l}-t-2l(t-\frac{3}{2})-1$\cite{Veron3}%
\end{tabular}
& $d_{2}=d_{1}^{\ast }$ \\ \hline
\begin{tabular}{l}
$\Gamma (L_{3},G_{3})$, where \\ 
$G_{3}(x)=A^{t}x^{t}+Ax+C$ \\ 
parity check matrix:%
$\left[ 
\begin{array}{cc}
H_{4}^{\ast } & 0 \\ 
1...1 & 1%
\end{array}%
\right] $
\end{tabular}
& $n_{3}=2^{2l}-t$ & $k_{3}=k_{2}$ (Lemma 3) & $d_{3}=d_{1}^{\ast }$ \\ 
\hline
\begin{tabular}{l}
$\Gamma (L_{3}^{\ast },G_{3})$, where \\ 
$G_{3}(x)=A^{t}x^{t}+Ax+C$ , $L_{3}^{\ast }=L_{3}\backslash \{0\}$ \\ 
parity check matrix:%
$\left[ 
\begin{array}{c}
H_{4}^{\ast } \\ 
1...1%
\end{array}%
\right] $%
\end{tabular}
& $n_{3}^{\ast }=2^{2l}-t-1$ & $k_{3}^{\ast }=k_{3}-1$ (Definition 2) & $%
d_{3}=d_{1}^{\ast }$ \\ \hline
\begin{tabular}{l}
$\Gamma (L_{4},G_{4})$, where \\ 
$G_{4}(x)=A^{t}x^{t}+A^{t-1}x^{t-1}+1$ \\ 
parity check matrix: $H_{4}^{\ast }$%
\end{tabular}
& $n_{4}=2^{2l}-t$ & $k_{4}=k_{4}^{\ast }$ (Corollary 2) & $d_{4}=d_{7}$ \\ \hline
\begin{tabular}{l}
$\Gamma (L_{4}^{\ast },G_{4})$ , where \\ 
$G_{4}(x)=A^{t}x^{t}+A^{t-1}x^{t-1}+1$ , $L_{4}^{\ast }=L_{4}\backslash \{0\}$ \\ 
parity check matrix: $H_{4}^{\ast }$%
\end{tabular}
& $n_{4}^{\ast }=2^{2l}-t-1$ & $k_{4}^{\ast }=k_{3}$ (Lemma 6) & $%
d_{4}^{\ast }=d_{7}$ \\ \hline
\begin{tabular}{l}
$\Gamma (L_{5},G_{5})$, where \\ 
$G_{5}(x)=Cx^{t+1}+A^{t}x^{t}+Ax$ \\ 
parity check matrix: $H_{4}^{\ast }$%
\end{tabular}
& $n_{5}=2^{2l}-t-1$ & $k_{5}=k_{4}$ (Lemma 7) & $d_{5}=d_{7}$ \\ \hline
\begin{tabular}{l}
$\Gamma (L_{6},G_{6})$, where \\ 
$G_{6}(x)=Rx^{t+1}+V^{t}x^{t}+Vx+1$ \\ 
parity check matrix: $H_{4}^{\ast }$%
\end{tabular}
& $n_{6}=2^{2l}-t-1$ & $k_{6}=k_{4}$ (Lemma 8) & $d_{6}=d_{7}$ \\ \hline
\begin{tabular}{l}
$\Gamma (L_{7},G_{7})$, where \\ 
$G_{7}(x)=x^{t+1}+1$ \\ 
parity check matrix: $H_{4}^{\ast }$%
\end{tabular}
& \multicolumn{1}{|l|}{$n_{7}=2^{2l}-t-1$} & \multicolumn{1}{|l|}{%
\begin{tabular}{l}
$k_{7}=k_{4}$ (Lemma 9) \\ 
$k_{7}=2^{2l}-t-2l(t-\frac{3}{2})-1$\cite{Veron2}%
\end{tabular}%
} & 
\begin{tabular}{l}
$d_{7}=2t+3$ \\ 
(Lemma 11)%
\end{tabular}
\\ \hline
\end{tabular}
\end{figure*}

In Table 2 we present the quasi-cyclic Goppa codes from our chain. 
It is easy to see that by substituting $x$ by $\beta x+\gamma $ we will obtain the same code if the Goppa polynomial is invariant to this substitution: $\alpha G(x)=G(\beta x+\gamma )$ \ where 
$\alpha,\beta,\gamma \in GF(2^{2l})$ .

In Table 2 we present such values of $\gamma $ and $\beta $ for the  Goppa codes from our chain.\smallskip 
 \begin{figure*}[!p]
\begin{center} 
Table 2 Quasi cyclic Goppa codes 
\end{center}
\begin{center} 
\begin{tabular}{|l|l|l|l|}
\hline
code & G(x) & $\gamma $ & $\beta $ \\ \hline
$\Gamma (L_{1},G_{1})$ & $x^{t-1}+1$ & $0$ & any nonzero element from $%
GF(2^{l})$ \\ \hline
$\Gamma (L_{2},G_{2})$ & $A^{t}x^{t}+Ax$ & $A^{-1}or$ $0$ & any nonzero
element from $GF(2^{l})$ \\ \hline
$\Gamma (L_{3},G_{3})$ & $A^{t}x^{t}+Ax+C$ & $\gamma \in GF(2^{2l}) :$ $A^{t}\gamma
^{t}+A\gamma =C(1+\beta )$ & any nonzero element from $GF(2^{l})$ \\ \hline
$\Gamma (L_{5},G_{5})$ & $Cx^{t+1}+A^{t}x^{t}+Ax$ & $\gamma \in GF(2^{2l}) :C\gamma
^{t+1}+A^{t}\gamma ^{t}+A\gamma =0$ & $\beta =\frac{\gamma C}{A^{t}}+1$ \\ 
\hline
$\Gamma (L_{7},G_{7})$ & $x^{t+1}+1$ & $0$ & $\left( \alpha
^{2^{l}-1}\right) ^{i}$ $,i=1,...,2^{l}+1$ \\ \hline
\end{tabular}
\end{center}

$\alpha $ is a primitive element of $GF(2^{2l})$ and $A\in GF(2^{2l}).$
%\caption{Quasi cyclic Goppa codes}
\end{figure*}

Therefore these quasi-cyclic codes have indexes $2^{l}-1$ and $2^{l}+1.$


\begin{thebibliography}{9}
\bibitem{Goppa1}V. D. Goppa, A new class of linear error correcting codes. Probl. Inform.Transm, Vol. 6, No. 3 , pp. 24-30,1970
 
\bibitem{Loel1}M.Loeloeian and J.Conan, A (55,16,19) binary Goppa code \emph{IEEE Trans. on Information Theory}, vol. 30, p.773, 1984.

\bibitem{Bezz5} S. V. Bezzateev, E. T. Mironchikov and  N. A. Shekhunova, One subclass of binary Goppa codes, Proc. XI Simp. Po Probl. Izbit. v Inform. Syst. pp. 140-141, 1986. 
  
\bibitem{Bezz1}S.V.Bezzateev and N.A.Shekhunova , On the designed distance of the best known (55,16,19) Goppa code , Probl.Inform.Transm., vol. 23, No 4, p.352, 1987.

\bibitem{Loel2} M.Loeloeian and J.Conan , A transform approach in Goppa
codes, \emph{IEEE Trans. on Information Theory}, vol. 35, pp.105-115, 1987.

\bibitem{Roseiro1} A.M.Roseiro, J.I.Hall, J.E.Adney and  M.Siegel, The trace operator and redundancy of Goppa codes, \emph{IEEE Trans. on Information Theory}, vol. 38,No.3, pp. 1130-1133, 1992.
 
\bibitem{Bezz2} S.Bezzateev and N. Shekhunova , Subclass of binary Goppa codes with minimal distance equal to the design distance, \emph{IEEE Trans. on Information Theory}, vol. 41, pp. 554-555, 1995.

\bibitem{Veron3} P. Veron, True dimension of some binary quadratic trace Goppa codes, Designs, Codes and Cryptography, 24, pp. 81-97, 2001.

\bibitem{Veron2} P. Veron, Proof of conjectures on the true dimension of some binary Goppa codes, Designs, Codes and Cryptography, 36, pp.317-325, 2005.

\bibitem{Bezz3}N.A.Shekhunova, S.V.Bezzateev  and E.T.Mironchikov,
A subclass of binary Goppa codes, Probl.Inform.Transm.,
vol.25, no.3, pp.98-102, 1989.

\bibitem{Bezz6}S.V.Bezzateev and N.A.Shekhunova, Quasi-cyclic Goppa codes, IEEE International Symposium on Information Theory, Canada, p.499, 1995.

\bibitem{Bom1} G. Bommer  and F. Blanchet, Binary quasicyclic Goppa
codes, Designs, Codes and Cryptography, 20,  pp.107-124,2000.

\bibitem{Veron1} P. Veron, Goppa codes and trace operator, \emph{IEEE Trans.
on Information Theory}, vol. 44,No.1, pp. 290-295, 1998.

\bibitem{Maatouk1} G. Maatouk, A.Shokrollahi and M.Cheraghchi,Good Ensembles of Goppa Codes, \emph{Ecole Polytachnique Federale De Lausanne,ALGO Lab}, , 2007, www. algo.epfl.ch/contents/output/sempr/Ghid\_MAATOUK.pdf

\bibitem{Bezz4} S.V.Bezzateev and N.A.Shekhunova ,  A subclass of binary
Goppa codes with improved estimation of the code dimension, Designs, Codes
and Cryptography, 14, pp.23-38, 1998
 
  
\end{thebibliography}
\end{document}